
\input harvmac

\Title{RU-94-92}{\vbox{The Cosmological Moduli Problem, Supersymmetry Breaking
, and Stability in Postinflationary Cosmology }}
\bigskip
\centerline{\it
T. Banks  \footnote{*}
{\rm  John S. Guggenheim Foundation Fellow, 1994-95;
Varon Visiting Professor at the Weizmann Institute of Science;
Supported in part by the Department of Energy under grant No.
DE-FG05-90ER40559  .} }
\smallskip
\centerline{M. Berkooz}
\smallskip
\centerline{Department of Physics and Astronomy}
\centerline{Rutgers University}
\centerline{Piscataway, NJ 08855-0849}
\smallskip
\centerline{\it P.J.Steinhardt\footnote{**}{\rm John S. Guggenheim
Fellow 1994-95, Supported in part by the Department of energy under
grant No. DOE-EY-76-C-02-3071, by National Science Foundation Grant
NSF PHY 92-45317, and by Dyson Visiting Professor Funds of the Institute
for Advanced Study}\footnote{$\dagger$}{\rm On Leave at the Institute for
Advanced Study, Princeton, NJ}}
\smallskip
\centerline{Department of Physics}
\centerline{University of Pennsylvania}
\centerline{Philadelphia, PA 19104}
\noindent
\bigskip
\baselineskip 18pt
\noindent
We review scenarios that have been proposed to solve the cosmological problem
caused by moduli in string theory, the Postmodern Polonyi Problem (PPP).  In
particular, we discuss the difficulties encountered
by the apparently \lq\lq trivial" solution of this problem,
in which moduli masses are assumed to
arise from nonperturbative, SUSY preserving, dynamics at a scale higher than
that
of SUSY breaking.
This suggests a powerful {\it cosmological vacuum selection principle}
in superstring theory.  However, we argue that if one eschews the possibility
of cancellations between different exponentials of the inverse string coupling,
the mechanism described above cannot stabilize the dilaton.  Thus, even if
supersymmetric dynamics gives mass to the other moduli in string theory,
the dilaton mass must be generated by SUSY breaking, and
dilaton domination of the energy density of the universe cannot be avoided.

We conclude that the only proposal for solving the PPP that works
is the intermediate scale inflation scenario of Randall and Thomas.
However, we point out that all extant models have ignored unavoidably large
inhomogeneities in the cosmological moduli density at very early times,
and speculate that the effects associated with nonlinear gravitational
collapse of these inhomogeneities may serve as an efficient mechanism
for converting moduli into ordinary matter.

As an important byproduct of this investigation, we show
 that in a postinflationary universe, minima of the effective potential with
negative cosmological constant
are not stationary points of the classical equations of scalar field cosmology.
Instead, such points lead to catastrophic gravitational collapse of that part
of
the universe which is attracted to them.   Thus postinflationary cosmology
dynamically chooses nonnegative values of the cosmological constant.
This implies that supersymmetry {\it must} be broken in
any sensible inflationary cosmology.
We suggest that further study of the cosmology of moduli will lead to
additional important insights about cosmology, SUSY breaking, and the
choice of the vacuum in superstring theory.

\Date{October 1994}

\newsec{Introduction}

The modular problem of string cosmology\ref\bkn{T.Banks, D.Kaplan, A.Nelson,
{\it Phys. Rev.}{\bf D49},779,(1994).}\ref\deCastro{B. de Castro, J.A.Casas,
F.Quevedo, E.Roulet, {\it Phys. Lett.}{\bf B318},447,(1993).} is a modern
version
of the cosmological difficulties created by the Polonyi field in the
earliest versions of spontaneously broken
supergravity\ref\polonyi{G.D.Coughlan, {\it et. al.},{\it Phys. Lett.}{\bf
B131},59,(1983).}.  One may call it
the Postmodern Polonyi Problem (PPP).
Briefly, in hidden sector models of SUSY breaking (with gravitational strength
forces playing the role of messenger\ref\hlw{L.J.Hall, J.Lykken, S.Weinberg,
{\it Phys. Rev.}{\bf D27},2359,(1983).}) there often exist
scalar fields with masses on the order of the weak scale and gravitational
strength coupling to ordinary matter.  Even in inflationary cosmologies
these fields behave like nonrelativistic matter just after inflation and
dominate the energy density of the universe until it is too low for
nucleosynthesis to occur.

In generic hidden sector supergravity models, such fields are needed to
generate
gaugino masses of order the weak scale.  One can eliminate them by
choosing a model in which SUSY is broken at low energies.  In string theory
we have no such luxury.   Massless moduli fields
exist in all known string ground states.  They parametrize the continuous
ground state degeneracies characteristic of supersymmetric theories.
Even if one were to find a ground state with no {\it geometrical} moduli\foot
{We use this phrase to describe moduli associated with the internal conformal
field theory of a string ground state which is the tensor product of four
flat spacetime dimensions and a conformal field theory with a discrete spectrum
of conformal dimensions.} one would still have the model independent dilaton
superfield.  If these moduli fields, which are massless to all orders in
perturbation theory, get their mass from the same nonperturbative mechanism
which breaks SUSY, and if the SUSY breaking $F$ term is $< 10^{10} - 10^{11}$
$GeV$
,as it is in all known models of SUSY breaking\foot{$\ldots$ with the possible
exception of models with continuous noncompact global symmetries that have
been discussed by Binetruy and Gaillard\ref\maryk{P.Binetruy, M.K.Gaillard,
{\it Nucl. Phys.}{\bf B358},121,(1991), {\it Phys. Lett.}{\bf B232},83,(1989),
{\it Phys. Lett.}{\bf B195},382,(1987).}.  These models cannot
be exact consequences of string theory, but it is possible that discrete
remnants of the noncompact symmetries used by these authors
are sufficient to obtain their results.}, then the moduli pose a cosmological
problem.

In this paper we will survey attempts that have been made to resolve this
problem.  In our opinion, no completely satisfactory resolution of the
cosmological moduli problem has been discovered, although the proposal of
Randall and Thomas seems promising.  Nonetheless, all
extant models require new and interesting phenomena to occur, both in
the realm of cosmology and in the theory of SUSY breaking.   We view this as
an indication that a correct theory of the cosmology of moduli will have
important, and probably testable, consequences.

In Section II of this paper we review existing proposals for solving the
cosmological moduli problem, including an unpublished (because unworkable)
proposal by Cohen, Nir, Moore and one of the present authors\ref\cohen{T.Banks,
A.Cohen, G.Moore, Y.Nir, {\it unpublished}.}.
The proposal of
Randall and Thomas\ref\rt{L.Randall, S.Thomas, MIT preprint, MIT-CTP 2331,
hep-ph 9407248.}
is reviewed.  It seems to successfully solve the PPP and can be made
technically natural by imposing a certain discrete R symmetry on their model.
Louis and Nir\ref\yossi{J.Louis, Y.Nir, Munich preprint, LMU-TPW-94-17, hep-ph
9411429.}
 have investigated
models which incorporate at tree level the mechanism proposed by Binetruy and
Gaillard \maryk .  They show that, in generic vacuum states, radiative
corrections drastically limit the separation that one can achieve between
the moduli masses and the weak scale.  The allowed separation is marginally
satisfactory from the cosmological point of view but all of these models
have flavor changing neutral currents that are too large to be compatible
with experiment.  We briefly discuss the question of whether there are
specific vacuum states in which radiative corrections to the Binetruy-
Gaillard mechanism are hierarchically small.

We then turn to proposals for giving mass to
the moduli at a scale higher than that of nonperturbative SUSY breaking.
This turns out to be more difficult than it sounds.
 In supergravity the manifold of chiral superfields
must be a Kahler manifold, and the
effective potential has the
form
\eqn\sugrapotex{V = e^K [D_i W D_{\bar{j}}\bar{W} K^{i\bar{j}} - 3\vert W
\vert^2 ]}
Here $K$ is the Kahler potential, $D_i$ the corresponding Kahler covariant
derivative, and $W$ the superpotential.
In order to have a supersymmetric ground state with vanishing cosmological
constant,
the superpotential and all of its first derivatives must vanish at the minimum
of the effective potential. This is a nongeneric condition, involving $n+1$
equations for $n$ unknowns.  It can be satisfied \lq\lq naturally'' if both
supersymmetry (which requires the Kahler derivatives of $W$ to vanish) and
a complex R symmetry (which requires the superpotential to vanish) are
preserved.
We argue that in the context of the conventional gaugino condensation
description of nonperturbative effects in string theory\foot{The arguments
are definitely more general than this, and probably apply to a wide
range of nonperturbatively generated superpotentials in SUSY gauge theories.}
the vanishing of the superpotential can only occur
 at points in moduli space where chiral multiplets charged under the hidden
sector gauge group become massless.  No such points have been found on
 any of the submanifolds of moduli space yet explored.

Thus, while there are
examples of nonperturbatively generated superpotentials which give rise
to stable, supersymmetric ground states, these ground states generically
break all R-symmetries.  The superpotential is nonvanishing at the
potential minimum.  As a consequence such a state will exhibit a large
negative cosmological constant.

In Section III, probably the most important section of this paper, we
show that a negative cosmological constant is more than just a phenomenological
embarrassment.  A universe that has undergone inflation cannot settle
in to a minimum of the potential with negative vacuum energy.  Instead,
it undergoes a violent recontraction on microscopic time scales.
Thus, if a system has several minima of its effective potential, some of which
have negative energy, cosmological evolution will favor those with nonnegative
energy.

The argument described above is based on the classical equations of
cosmological evolution.  States with nonnegative vacuum energy are
potentially unstable to decay into negative energy states via quantum
tunneling.  We show that in theories of moduli, and more generally in
"natural" models of inflation, the tunneling amplitudes are
less than $e^{- 10^{12}}$ per unit space time volume,
 and might be identically zero.  A zero energy ground state
in such a theory has a lifetime for decay into hypothetical negative energy
states, which is much longer than the age of the universe. This implies that
even though our own universe might be unstable in such a model,
we have no need to
worry about living in such an unstable world.

In Section IV we return to an examination of supersymmetric ground states
which can freeze the moduli in string theory.  We argue that the
restriction to states with vanishing cosmological constant is indeed
very strong in this context.  As mentioned above,
 the requirement appears to force
us to sit at a point in moduli space where extra chiral multiplets charged
under the hidden sector gauge group, become massless.  There are no known
points where this occurs.  We discuss the implications, and provide
a favorable interpretation, of this negative result.  It suggests that
the search for supersymmetric ground states with nonperturbatively vanishing
cosmological constant may lead to isolated points in string moduli space.
Thus, the cosmological selection of such states becomes a dynamical
vacuum selection principle for string theory.

We point out
that the mechanism under discussion can probably not give mass to the
dilaton.  Finally,
we examine the generation of a dilaton mass by SUSY violating
phenomena, and confirm that all known mechanisms for SUSY breaking
still lead to cosmological disaster.

In passing, we provide a mechanism
for cancellation of the cosmological constant in scenarios of low energy
SUSY breaking.  We show that it requires the existence of a light weakly
coupled field (the dilaton in our case) with a mass of order $10^{-2} -
10^{-3}$
$eV$.  In the presence of such a field, the cancellation of the
cosmological constant in theories with low energy SUSY breaking is no
more unnatural than it is in hidden sector models.
In addition, the dynamics of a field with such a small mass
 might conceivably explain the fine tuning of the cosmological
constant to levels consistent with observation.  Unfortunately, it is
precisely this light field which dominates the energy density of the
universe in these models, leading to a PPP.  In this case, the ideas of
\cohen can be made to work, saving nucleosynthesis.  However, the ratio
of nonrelativistic matter to radiation in the present era is
predicted to be many orders of magnitude larger than it actually is.

In section V we present a highly speculative scenario which might
resolve the cosmological moduli problem in a novel way.  We point out
that in all extant models, moduli do dominate the energy density of the
universe for a long time after inflation.  Since they behave like
nonrelativistic matter, inhomogeneities in the moduli fields grow with
the expansion.  We show that they go nonlinear long before the moduli
decay.  This leads us to
speculate that nonlinear processes associated with gravitational
collapse ({\it e.g.} the formation of stable, gravitationally bound
{\it modular stars}) could lead to an
enhancement in the decay of moduli into ordinary matter, thus
eliminating the moduli before the era of nucleosynthesis.
At present, we do not know how to calculate in this complicated
nonlinear regime, so we cannot assess
the viability of this proposal.

Finally let us note the recent paper of Bento and Bertolami, which also
treats the Polonyi problem in string theory \ref\bento{M.C.Bento, O.Bertolami,
gr-qc 9409059}
\newsec{Some Modest Proposals}

The general argument that moduli fields dominate the energy density of
the universe has a number of loopholes, and proposals to avoid the problem
have tried to exploit most of them.

\subsec{Intermediate Scale Inflation as a Solution to the PPP?}

We begin by reviewing the work of Randall and Thomas (RT)\rt who suggested
inflation with a weak scale Hubble parameter as a mechanism for diluting
the moduli.  If the Hubble parameter is of the order of the modular masses
then their energy density indeed redshifts away exponentially during inflation.
Randall and Thomas estimate that $7$ to $10$ e-foldings are sufficient
to reduce the modular energy density to an acceptable level.
They call their proposal weak scale inflation, but we prefer the name
Intermediate Scale Inflation (ISI),
because the intermediate scale $\sqrt{M_W M_P}$
is the fourth root of the inflationary vacuum energy in this model.

RT note that intermediate scale inflation produces density fluctuations
many orders of magnitude smaller than those required by
observations of inhomogeneities in the cosmic microwave background.
To resolve this, they invoke a previous era of inflation, with a higher
vacuum energy density.  They claim that the requirement that the primordial
fluctuations responsible for the observed microwave background distribution
not be blown up larger than our horizon volume by the second stage of
inflation, is
that there be fewer than $25 - 30$ e-foldings of ISI.
This is compatible with the amount of inflation necessary to eliminate
the moduli.

Unfortunately, the RT proposal appears to have a naturalness problem.
Indeed, such proposals were considered and telegraphically dismissed
by A. Nelson in a cryptic footnote in \bkn .  The problem is
that in order to obtain sufficient reheating (i.e. in
order for the Randall-Thomas inflaton {\it not} to
pose the same kind of problem as the moduli),
one must couple the inflaton to ordinary matter via renormalizable couplings.
Generically, this would lead to a renormalized effective potential for
the inflaton field which varied when the inflaton field changes by an
amount of order the weak scale.  Such a potential cannot lead to inflation.

Randall and Thomas propose to deal with this problem by invoking SUSY
nonrenormalization theorems.  They would like to have a term in the
superpotential coupling the inflaton, $I$ to low energy fields
in a renormalizable way, but no terms in the superpotential that
depend only on $I$. Neither can they afford to have soft SUSY breaking
terms containing only a single power of the SUSY breaking F-term
multiplied by a function of ${I\over M_P}$.  Any superpotential
term or soft SUSY breaking term of this type, will generate
a potential much larger than the fourth power of the intermediate
scale.  If no such terms occur, then the potential for $I$ can vanish
up to linear order in $F$, at the minimum for the other fields.

The solution they find to this infinite set of conditions is clever and
probably unique.  A discrete R symmetry under which $I$ is neutral
forbids all the relevant terms.  In addition one must assume no elementary
fields, $\Phi_2$ with R charge $2$, to prevent the appearance of
terms of the form $\Phi_2 G({I\over M_P})$ in the superpotential.
Finally one must worry about the necessary breaking of the R-symmetry.

Surprisingly, this does not pose any problems.  Generic nonrenormalizable
hidden sector (NRHS) \bkn models contain a strongly interacting sector
which breaks all complex R symmetries at a scale $M_R$ related to the SUSY
breaking scale by $M_R \sim (F M_P)^{1\over 3}$.  This is necessary
in order for the cosmological constant to be zero.  The field whose
F term breaks SUSY carries R charge zero (and so its F term carries R charge
-2) because it is a flat direction of the tree level potential, whose VEV
is being fixed by nonperturbative physics at scale $M_R$.  The R symmetry
is a valid symmetry along this flat direction only if the superfield
carries R charge zero.

In any such model, all R breaking terms in the action of the field $I$
will have their origin in its coupling to the strongly interacting sector
at $M_R$.  If all such couplings are nonrenormalizable (which is
automatically the case if the strongly coupled theory is a pure SUSY gauge
theory or one with matter in purely chiral representations), the R breaking
superpotential induced for $I$ will have the form $M_R^3 w({I\over M_P})$
and will give an intermediate scale potential (remember that in such theories
$M_I^2 \sim {M_R^3 \over M_P}$).

Renormalization of the Kahler potential of $I$ by weak scale loops will
modify this potential in the regime where $I$ is much smaller than
the Planck scale and inflation will not occur when $I$ is in this regime.
However, the asymptotic behavior of these corrections for large $I$
will be of the form
\eqn\modifpot{\delta V \sim {\lambda^2 \over 8\pi^2} ln({I^2 \over M_P^2}) V}
where $\lambda$ is the renormalizable Yukawa coupling of $I$ to weak scale
matter.  For a reasonable range of small values of $\lambda$, these
corrections are negligible compared to $V$ itself, even when $I$ is
a few times the Planck scale.

Thus, the imposition of a discrete R symmetry makes the Randall-Thomas
proposal technically natural within the framework of NRHS theories.
It remains to be seen if one can actually find a string vacuum state
with a field with the properties of $I$.
One must further study the effect on this proposal
of the early gravitational clumping of
moduli that we will discuss in the final section.
With these caveats in mind though, one can conclude that
Intermediate Scale Inflation is an acceptable solution of the PPP.
It remains to be seen whether such a model can be derived from string theory
and whether the double inflation scenario invoked by Randall and Thomas
truly emerges naturally from the dynamics of a specific model.

\subsec{Can One Have a Hierarchy Between Squark and Moduli Masses?}

A second loophole in the argument that moduli dominate the energy density
of the universe is the assumption that  the highest scale of SUSY breaking
$F$, is related to the squark masses by $F \leq M_P M_{sq}$.  The motivation
for this assumption is the fact that one can write dimension $6$
operators invariant under any compact symmetry which couple the superfield
whose $F$ term breaks SUSY, to the quark superfields.  These terms
generate squark masses when SUSY is broken.  It is reasonable to
assume that the relevant scale for these nonrenormalizable couplings
is not larger than the Planck mass, whence the bound.
Binetruy and Gaillard \maryk have suggested that noncompact global symmetries
can suppress squark masses by higher powers of the Planck scale.
The continuous global symmetries that they invoke cannot be
exact symmetries of string theory, but one might hope that
some discrete remnant of these symmetries could do the job.

One cannot hope for such a mechanism to be stable under radiative
corrections in a theory in which SUSY solves the hierarchy problem.
Stability of scalar masses under gravitational radiative corrections
bounds the gravitino mass by about $10^{11}\quad GeV$.  Louis
and Nir \yossi have studied the one loop radiative corrections to vanishing
scalar masses around conventional string vacua.  They find that
generically, the one loop corrections due to light fields with
gauge charge are nonvanishing.  Thus, in the vacuum states they
studied, the inequality relating the high scale of SUSY breaking
to the squark masses can only be weakened by a single factor of
$\alpha_3 \over \pi$ .  This may allow us to raise the moduli
masses high enough to allow for nucleosynthesis.  It is certainly
not enough to allow the temperature to which the universe is reheated
after modular decay to be high
enough to ignite weak scale baryogenesis.  We will see below that
this may not be too much of a problem.

Unfortunately, Louis and Nir also showed that models of this type
have large flavor changing neutral currents as a consequence of
string loop corrections.  They are unlikely to be compatible with
experiment.  We note however that Louis and Nir studied generic
vacuum states which have noncompact continuous symmetries at tree
level.  We know that these symmetries are broken by loop corrections,
but some discrete noncompact subgroup might be preserved in {\it particular
vacuum states}.  It is conceivable that in such special vacuum states
the radiative corrections to squark masses are hierarchically small,
as suggested by Binetruy and Gaillard.  Perhaps the flavor problems
are also mitigated in vacuum states invariant under noncompact discrete
groups.

\subsec{Saviours of the Universe?}

We now turn to the proposal of \cohen for solving the reheating
problem.  Although unsuccessful, it illustrates some
interesting features that may reappear in a more robust theory of moduli.
The value we have been using for moduli masses is an order of
magnitude estimate.  Let us assume that for one or more of the moduli
fields, this estimate is off by the rather large factor of $30$.
Then some of the moduli will have a reheat temperature\foot{We use
the phrase {\it reheat temperature of the moduli} to refer
to the temperature of the gas of light particles produced by thermalization
of the products of modular decay.}  of order
a few $MeV$, hot enough for nucleosynthesis.\foot{The proposal to
solve the Polonyi problem by raising the Polonyi mass above the weak scale
was apparently first made by the authors of\ref\ellis{J.Ellis,
D.V.Nanopoulos, M.Quiros, {\it Phys. Lett.}{\bf B174},176,(1986)}.
Recently Yamaguchi {\it et. al.} investigated a similar suggestion for
solving the Polonyi problem in a more general context\ref\yamaguchi{
M.Yamaguchi, T.Yanagida, T.Moroi, Tohoku preprint TU-467 hep-ph 9409367.}.}.
More importantly,
until the energy density falls to this value, the heavy moduli
(which we will dub the {\it saviors}) behave just like the light
ones.  The ratio of energy densities in heavy and light moduli
is of order one\foot{Actually it is more like the ratio of
the number of fields of each type.  This leads one to search
for string ground states with a small total number of moduli,
something that may be called {\it the minimum modulus principle}.}
at the end of inflation, and remains constant until the saviors
decay.  Thus, if the initial ratio is somewhat larger than one,
we will have a radiation dominated universe at the energy density
relevant for nucleosynthesis.

Of course, the temperature will never be high enough for even weak scale
baryogenesis, but this is unnecessary.  All of the moduli have only
Planck scale couplings to ordinary matter, and it is perfectly
consistent with all data on baryon number conservation to
assume that these couplings violate baryon number and CP.
Thus, baryogenesis could arise from the (obviously out of
equilibrium ) decay of the saviors\foot{Making them the
creators of all matter as well as its saviors.}.
The baryon to photon ratio produced in this decay process would
be the inverse savior mass, measured in $MeV$, times the asymmetry
in a given decay.  Thus, a factor of $10^{-7} - 10^{-8}$ in the
baryon to entropy ratio just represents the small ratio between the
reheat temperature and the mass of the saviors.  The rest of
the observed suppression of the baryon to photon ratio
could come from a weak coupling factor.  The decay of the heavy
savior into conventional matter can be computed in the parton model
in terms of the matrix elements of the leading operator which causes the
decay.  In order to see a CP violating phase and obtain an asymmetry
one must interfere tree level and higher order diagrams
and pay the price of a loop factor.  It is not implausible then
that such a model could reproduce the observed baryon asymmetry.

The problem of this model comes with the decay of the light moduli.
Although the entropy produced in their decay does not wash out
the baryon asymmetry, the details of the decay process completely
change the element abundances produced in the (presumed successful)
nucleosynthesis that followed savior decay.
Moduli are very heavy, and their decays will produce hard photons
and hard hadron jets.  These will thermalize their energy, initially
through hadronic collisions, but in the process will first
produce large numbers of photons capable of disintegrating deuterium .
Dimopoulos {\it et. al.}\ref\hall{S.Dimopoulos, R.Esmailzadeh, L.J.Hall,
G.D.Starkman, {\it Nucl. Phys.}{\bf B311},699,(1989). }
 have studied this problem in great detail for gravitinos,
and have come to the conclusion that the fraction of energy density in
heavy decaying particles in such a situation cannot be larger than $10^{-6}$.
In the model of \cohen we cannot reasonably expect this fraction
to be smaller than a tenth.

To summarize, although several ideas have been proposed for resolving
the cosmological moduli problem, only Intermediate Scale Inflation
appears to hold out any promise.
The models discussed so far retained the assumption that the
physics responsible for moduli masses was also the agent of
dynamical SUSY breaking.  At first sight, the most reasonable
resolution of the whole problem would seem to be decoupling these
two nonperturbative effects.  Moduli get their mass from
dynamics at a higher scale than SUSY breaking.  This seems particularly
plausible in view of the fact that we have two hints
in the present data of the existence of a new scale of physics
at $10^{16} - 10^{17} ~GeV$.  These are the ``observed''
unification of couplings, and the vacuum energy density
required to explain the COBE microwave background anisotropy data
in inflationary cosmology.  If physics at such a scale
generated moduli masses, the moduli would decay long before the
beginning of the classical period of cosmic history.

It is somewhat surprising to find that this simple solution does
not really work.  More precisely, we show below that the nonperturbatively
generated superpotential for moduli must vanish
in the vacuum in order to have an acceptable cosmology.
This nongeneric condition is not satisfied by any know superstring
vacuum.  Further, we argue that a SUSY preserving superpotential
cannot stabilize the
dilaton unless we are willing to imagine the cancellation of
two effects which are of different order in the weak coupling expansion.

Before proceeding to demonstrate these facts, we must pause for
an act of iconoclasm.  Our discussion will hinge on the fact that
typical supersymmetric minima of the potential have negative vacuum energy.
Everyone would agree that this is not good for phenomenology.
What we will demonstrate is that in the context of inflationary
cosmology, such minima do not even correspond to stationary states
of the system.  This will lead to a powerful cosmological selection
principle for superstring vacua.

\newsec{The Importance of Being Nonnegative}

{\it All systems seek their state of lowest energy,} is a maxim that
physicists learn sometime in their preschool years.  Like most
convenient aphorisms, it summarizes the behavior of an often complicated
set of rules in a way that is easy to remember and easy to
apply.  The utility of this principle in physics has been so great that
it is somewhat shocking to find that there are systems, such as spin
glasses, to which it does not apply.  It is little wonder then that most
discussions of fundamental cosmology assume that the universe is tending
towards a stable vacuum state as time goes on, and that this state has
the lowest energy density allowed by the basic lagrangian.

Andrei Linde pointed out long ago that observations do not require such
absolute stability\ref\linde{A.D.Linde, {\it Phys. Lett.}{\bf B70},306,(1977).
}.  If one wants to place a rigorous
theoretical lower bound on
the mass of the Higgs boson by requiring that the standard model vacuum
be stable, the most one can honestly ask for is that its lifetime be
longer than the observed ``age of the Universe''\foot{More precisely,
``the age of our current horizon volume according to conventional cosmology''.}
Linde's calculations referred to a tunneling instability, in which weak
couplings can easily explain a very long lifetime.

Despite these observations, there has been a certain amount of unease in
the astroparticle community about the prospect that the state of the
universe in which we find ourselves is not absolutely stable.  Perhaps
the strongest concrete reason for this unease has to do with the
genericity of initial conditions.  We know that our universe once had a
much larger energy density than it does today.  If the barrier that
prevents the classical decay of our hypothetically false vacuum was
smaller than the available energy density in thermal motions {\it at any
time in the past history of the universe.}, it is very hard to
understand how we came to rest at this false minimum.  The universe
should ``overshoot'' the false vacuum and end up in the true minimum of
the potential.  This expectation is based on the maxim with which we began
this section.

The main thing that we wish to demonstrate in this section is that this
argument is false, if we assume certain natural conditions:

\noindent
1.The present history of the universe was preceded by a period of
inflation, during which all spatial curvatures and scalar field
inhomogeneities were washed out.

\noindent
2.The cosmological constant in the hypothetical false vacuum in which we
live is zero.  Our results are not changed very much if we assume a
positive cosmological constant.

Assume that the scalar field potential of our model has a locally stable
minimum with zero cosmological constant and one or more minima with
negative cosmological constant.  Then we will show below that the only
stable fixed point of the equations of motion is one in which the scalar
field sits at its zero energy minimum and the universe is flat and
static.  Solutions with generic initial conditions are not attracted
to this fixed point.  Instead, they lead to a situation in which the
universe contracts irreversibly (we assume it is initially expanding).
No solution comes to rest in the minimum energy density state.
This means that if we assume some probability distribution for the
initial conditions, the only solutions which lead to a stable evolution of a
large universe are those which asymptote to the false vacuum\foot{As a
consequence of the friction in an expanding universe, this is a set of
nonzero measure in the space of all initial conditions, with respect to
a flat probability distribution.}.  The
criterion for classical stability
in a post-inflationary expanding universe is very
different than that in flat space.

Our observations do not change the considerations of Coleman and DeLucia
about the quantum mechanical instability of the false vacuum.  However,
if we restrict attention to the class of models of modular dynamics
which we have been studying, then the tunneling amplitudes are very small.
In these models the dynamics of the early universe is dominated by a set
of scalar fields which have only string scale nonrenormalizable
couplings in the fundamental lagrangian and to all orders in
perturbation theory.  At an exponentially lower scale, $M$,
nonperturbative dynamics gives these fields a potential of order
 ${M^6 \over M_P^2}$.

In this general class of models, a simple scaling argument shows that
the instanton action which
controls the instability of the false vacuum is of order ${({M_P \over
M})}^6$, where $M_P$ is the string scale.
Thus for all reasonable values of $M$\foot{It seems to us that the maximum
reasonable value for $M$ is that which leads to the vacuum energy scale
($10^{16} - 10^{17} ~GeV$) required by the inflationary explanation of the
observed fluctuations in the microwave background. This
gives a tunneling amplitude
of order $e^{-10^{12}}$ per unit Planck volume.}the lifetime of the
false vacuum is enormously longer than the age of our horizon
volume.

\subsec{Scalar Fields in a Postinflationary Era}

The equations of motion for a set of scalar
fields in a post-inflationary cosmology with Robertson Walker scale
factor $R$, are
\eqn\scalar{{d\over dt}(M_P G_{ij}({z}) \dot{z^i}) + 3H M_P
G_{ij}(z)\dot{z^j} +{1\over M_P}{\partial V \over \partial z^i}
=0}
\eqn\H{H^2 \equiv ({\dot{R} \over R})^2 = ({1\over M_P^2}) [\ha M_P^2 (G_{ij}
\dot{z^i}\dot{z^j} ) + V]}
Here $z^i \equiv {\phi^i \over M_P}$ are dimensionless scalar fields,
and $G_{ij}$ is the metric on the space of fields.
Note the absence of a spatial curvature term in Einstein's equation,
which signals that we have undergone a period of inflation. From
the second of these equations it follows immediately that
there are no solutions with $\phi_i$ coming to rest at a minimum with
negative value of the potential.  What then is the typical behavior if
the potential has a local zero energy minimum and a global minimum with
negative energy?

Combining the two equations we obtain
\eqn\enloss{\dot{E}\equiv \dot{H^2} = 2 H\dot{H} = -3 H \sum \dot{\phi_i}^2}
 In an expanding universe, $H$ is positive.  Equation \enloss says that
it decreases in magnitude as the universe expands.  What happens when
$H$ reaches zero?  Equation \H says that this can only happen at a place
where the potential is less than or equal to zero.  In particular, $H$
vanishes if we are sitting at rest in the zero energy false vacuum.  It
is easy to see that this is an exact solution to the equations of
motion.  Since, in an expanding universe,
 we are dealing with a system which has {\it friction}, a finite volume
 in phase space will be attracted to this fixed point.  Generically,
 solutions with initial conditions outside this volume will stop
 expanding when $H$ hits zero, as long as any field velocity is
 nonvanishing. Equation \enloss shows that $H$ will then change sign and
 the universe will begin to contract.  $H^2$ increases without
 bound, and the universe rapidly contracts back to
a singular state.  In effect, for initial conditions which lie in the basin
of attraction of the negative energy minimum, inflation is never successfully
completed.  If different regions of the universe exit inflation
with different initial conditions for the scalar fields, only those
regions which are attracted to the zero energy minimum will remain large
for any substantial period of time.

There are solutions
 with finely tuned initial conditions for which all velocities vanish at
 a nonstationary point on the potential surface with $V = 0$.
The system then enters a limit cycle, moving on the negative part of the
potential surface, with total energy equal to zero.  This limit cycle is
an exact solution of the full equations of the system, but an unstable
one.  Generic small perturbations of it drive one to a collapsing state.

We see that in a post inflationary universe, minimizing the energy is
not the way to achieve classical stability.
It should be emphasized that there is nothing in this argument that
depends strongly on the fact that the energy in the false vacuum is
zero. If it is positive there will be an attractive ``fixed point''
solution in which the fields are at rest in the false vacuum and
the metric is DeSitter.  Our arguments explain the classical stability
of a flat vacuum state relative to one with negative cosmological
constant, but do not provide any obvious clue as to why the cosmological
constant must vanish.

The foregoing line of reasoning does not imply the quantum mechanical
stability of the false vacuum state.  Indeed, decay of the false vacuum
proceeds by quantum tunneling in a finite sized bubble.  A prior era of
inflation does nothing to erase the spatial curvature inside this
bubble, and a local, anti-DeSitter metric is still the proper geometry
of the bubble interior after its formation.  The analysis of Coleman and
DeLuccia\ref\coleman{S.Coleman, F.DeLucia, {\it Phys. Rev.}{\bf D21}, 2133,
(1986).} remains valid, and their generically gloomy prediction
of the consequent fate of the world is unaltered.

The probability of
tunneling is of course highly model dependent.   If we restrict attention
to lagrangians for string theory moduli, the
 characteristic form of the effective potential is:
\eqn\effpot{V_{eff} = {M^6 \over M_P^2} V({z^i})}
where $M_P$ is the string scale.  The fields $z^i$ are assumed to be
canonically normalized at $z^i = 0$.  At other points in field space
they have a kinetic term
\eqn\kinterm{{\cal L}_{kin} = \ha M_P^2 G_{ij}({z^i})\nabla z^i
\nabla z^j}
 These formulae follow from the fact
that the potential is generated by a strongly interacting
supersymmetric gauge theory with scale $M$ and the moduli are
coupled to the strongly interacting fields only through
irrelevant operators scaled by the Planck mass.

Assume now that the effective potential has a zero energy minimum at
$z^i = 0$, and
another one with negative energy.  If we assume that the potential
contains no particularly large or small dimensionless constants, the
second minimum is at ${z^i} = o(1)$ and the negative vacuum
energy of order ${M^6 \over M_P^2}$.

The equation for the flat space instanton which controls the tunneling
rate from the false to true vacuum is
\eqn\inst{\nabla^2 (G_{ij}({z}){z^j}) = - {M^6 \over M_P^4} V_i ({z})}
The solutions will have the form
\eqn\soln{z^i = f^i \left( x {M^3 \over M_P^4}\right)}
where the functions $f^i (x)$ do not depend on either $M$ or $M_P$.
As a consequence, it is easy to see that the instanton action is of
order:
\eqn\action{S_{inst} = o({M_P \over M})^6}

In the class of models in which $M$ is the scale of nonperturbative
physics which is responsible for inflation, ${M_P \over M} \geq 10^{5\over 3}$
and the vacuum tunneling
probability per unit space time volume is of order
\eqn\tunprob{P \sim {M_P^{16} \over M^{12}} e^{- ({M_P \over M})^6} \leq
10^{24} e^{- 10^{10}} M_P^4}
This is so small that it is of no conceivable relevance to physics as
measured by local observers in our universe.  The
false vacuum is essentially stable.
Moduli dominated cosmological models that undergo inflation
can thus live happily in a false vacuum, with no fear of instability.

In fact, it is possible that the false vacua in these models are absolutely
stable.  Coleman and DeLucia
 showed that in many cases, no instanton exists
for tunneling from a space with nonnegative cosmological constant into
Anti-DeSitter space\ref\weincvet{S.Weinberg, {\it Phys. Rev. Lett.}{\bf 48},
(1982),1776; M.Cvetic, S.Griffies, S-J.Rey,{\it Nucl. Phys.}{\bf B389},
(1993),3.}
.  Heuristically this occurs because the spatial sections
of Anti-DeSitter space have constant negative curvature.  Thus, an
Anti-DeSitter bubble has constant surface to volume ratio as it grows,
and energy balance does not automatically induce the growth of large enough
bubbles.

It is easy to see that even in the presence of gravity,
all small parameters
scale out of the equations of motion which determine the existence of
instanton solutions.  Thus, the existence of instantons depends on
dimensionless numerical constants in the lagrangian.  It is conceivable that
in the modular lagrangians determined by string theory these constants
are such that no instanton exists, and the zero energy vacuum is
absolutely stable.  We emphasize however that this hypothetical possibility
is in no way necessary to our argument.  Even if modular instantons
exist, the tunneling rates which they predict are too small to be of interest
on time scales of order the age of our universe.

We believe that the instability of negative energy minima of the potential
in postinflationary cosmology is an important clue about the nature of
the universe.  In the next section we will show that it can be the basis of
a powerful vacuum selection criterion in superstring theory.

\newsec{Massless Multiplets and Massive Moduli}

The results of
the previous section put strong constraints on the idea that supersymmetric non
perturbative dynamics gives mass to the moduli.  Mass generation
for the moduli can only occur if a nonperturbative superpotential is
generated.  If SUSY is preserved, the stationary point of the
supergravity potential will have a negative cosmological constant of the
form $ - {3\over M_P^2}\vert W_{min}\vert^2$.  We have seen that fields
lying at rest in the minimum of the potential
will not be a stationary solution of the postinflationary
cosmological equations unless
$W_{min} = 0$.  We can phrase this important result in terms of symmetries:
{\it In postinflationary cosmology, breakdown of R-symmetry\foot{
To be precise, of any R symmetry larger than a $Z_2$ R-parity. We remind
the reader that any R-symmetry larger than $Z_2$ implies the
vanishing of the superpotential at a symmetric minimum of the potential.}
implies
that SUSY must be broken in a stable (Minkowski or DeSitter) vacuum state}.

Now consider the form of the nonperturbatively generated superpotential in
a string theory ground state.  We will assume for simplicity that at generic
points in moduli space, the strongly coupled gauge theory which generates
the superpotential is a pure, $N = 1$,
 SUSY gauge theory.  We believe that our results
are more general than this, but it would take us too far afield to delineate
the precise class of theories for which our discussion is valid.  For pure
gauge theory, the exact form of the superpotential is known to be
\eqn\puresupot{W_{NP} = M_P^4 \quad e^{- b S - \Pi (\Phi_i )}.}
Here, $S (= {8\pi^2\over g^2} + i\theta )$ is the dilaton superfield,
$\Phi_i$ are
the moduli, and $b$ is related to the one loop $\beta$ function for generic
values of the moduli.  $\Pi$ is the moduli dependent one loop renormalization
of the coupling coming from (generically) massive modes. At nonzero values of
$g$, \puresupot will only vanish if,
at some point in moduli space, $\Pi$ diverges.  This means that extra massless
fields appear at this point.  If we want to reduce the value of the
superpotential,
then these must be matter fields rather than gauge fields.  Indeed, the
example of SUSY QCD shows us that an increase in the number of
massless chiral multiplets
can lead to vanishing of the nonperturbative superpotential.  This theory
has a nonperturbative superpotential for $N_C > N_F$, but not for $N_F \geq
 N_C$.
We will assume that there are points in string moduli space
where such an increase in the number of matter multiplets in the hidden
sector occurs.  We do not know of an example of such a point, but, as we will
see, this fact may be interpreted in a positive manner.

At these points, $\Phi^0$, the superpotential will vanish like a power of
$\Phi_i - \Phi_i^0$ ($\Pi$ will blow up logarithmically as a consequence of the
existence of new massless states).  If this power is greater than one,
then $\partial_{\Phi_i} W$
will also vanish and SUSY will be preserved.
Under these conditions, string theory will have a locally
stable supersymmetric ground
state. (Of
course, SUSY can be broken by dynamics at a lower scale.)
If the point in moduli space where extra massless nonsinglet
chiral multiplets appear is isolated, then the fluctuations in all of the
moduli fields apart from the dilaton will be massive, with masses of order
$e^{-{bS}}M_P$. It would appear desirable then that the special points where
massless chiral multiplets are present, are dimension zero submanifolds of
moduli space, so that we can give mass to all the moduli.
This might explain why a survey of known ground states, which
explores a submanifold of moduli space with large codimension, fails to
reveal such a point\foot{We are well aware that this argument is at best
a good excuse.  The program that we are outlining requires us to at least
prove the existence of these special points in moduli space. We have so far
failed to do so.}.

The paragraph above is the promised vacuum selection principle for
superstrings.  Four dimensional classical ground states for string theory,
with nonabelian low energy gauge fields, will often lead to
nonperturbative
superpotentials for the moduli fields, and to spontaneous violation of
all complex R-symmetries.  Supersymmetric "ground" states
will generically have negative vacuum energy and will as we have seen, not
be stationary states of a postinflationary universe.  The hypothetical
points in moduli space where the superpotential vanishes will be dynamically
chosen by the equations of cosmology.

The above discussion assumed that SUSY was not spontaneously broken by
the nonperturbative dynamics.
We note in passing that the above arguments may also provide a hint of
the explanation of why SUSY is broken in the real world.  The
minimization of the potential and the vanishing of the superpotential
are $n+1$ complex equations for $n$ unknowns.  It may be that for the
full low energy superpotential, they have no common solution.  This
would mean that R-symmetry is spontaneously broken.  A successful
inflationary cosmology would then require that SUSY be broken as well.

As an example of how this could happen, imagine a class of string ground
states for which the low energy hidden sector gauge theory
was (at generic points in moduli space) a pure gauge theory with
a product of two groups $SU(N)$ and $SU(L)$ with $N > L$.
Suppose that, for $SU(N))$, there exists
an isolated
``magical'' point in moduli space,
at which $N_F > N$ massless chiral multiplets in the ${\bf N + \bar{N}}$
appear.
Assume that these are singlets under the second factor in the gauge group.
The argument above implies that the theory will have a supersymmetric
ground state at this point, and that it will be cosmologically chosen
in preference to other possible supersymmetric minima of the potential
for moduli.

Apart from the dilaton, the moduli will all be massive at this point.
Thus, below the first ``confinement'' scale, the low energy theory will
consist of a dilaton, a pure SUSY $SU(L)$ gauge theory,
and some massless fields
associated with the chiral symmetries of the massless matter in the
first factor of the gauge group.  The superpotential for these $SU(L)$
singlet fields will have a supersymmetric minimum, at which the
superpotential vanishes.  Gaugino condensation in $SU(L)$ will now
generate a superpotential $e^{- {S\over L}}$ for the dilaton below the $SU(L)$
confinement scale.  The structure of the dilaton potential
will be
\eqn\dilpot{V = e^{K - {(S + \bar{S})\over L}} [K_{S\bar{S}}^{-1}\vert {1\over
L} - K_S \vert^2 - 3]}
In the present state of our knowledge of string theory we can give only the
first term, $- ln(S + \bar{S})$ of the large $S$ asymptotic expansion of the
Kahler potential $K$.  Thus we can only speculate about the existence of
possible stable cosmological solutions of the equations of gravity coupled to
the dilaton in this model.  However, we can be sure that no such solution is
supersymmetric.  Hypothetical supersymmetric vacua at finite $S$ will have
negative vacuum energy and suffer from the instability described above.  The
state
at infinite $S$ suffers from the Dine-Seiberg instability.  Any stable solution
of this system violates SUSY.

This example teaches us another lesson about our attempt to freeze the moduli.
Even if our hypothetical points in moduli space exist, the superpotential given
above cannot completely finish the job of giving mass to the moduli at a
supersymmetric minimum, for
the superpotential vanishes for all values of the dilaton field if it vanishes
at all.  This result is quite general and could only be averted if the
superpotential had a complicated dependence on the dilaton, which seems
unlikely in the weak coupling region\foot{Many popular models of the dilaton
potential assume cancellations between different exponentially
small terms in the string coupling.  While no definitive argument that
such models are incorrect exists, we find them unpalatable.}.
The dilaton mass
must come from SUSY breaking dynamics, and we are not yet out of the
modular woods\foot{Note that, during inflation when $W_0$ is nonzero,
the dilaton may be quickly
driven to a "minimum" of the potential $V_I = W_0 e^{K - {(S + S ^*)\over N}}
[K^{S S^*}\vert -{1\over N} + \partial_S K \vert^2
+ K^{ij*}\partial_i (ln W_0 + K) \partial_{j*} (ln W_0^* + K) -
3]$\ref\drt{M.Dine, L.Randall, S.Thomas {\it Paper in preparation.}}.
This is unlikely to be the true minimum of the full dilaton
potential and so the dilaton
will generally start out its postinflationary motion a distance
of order $M_P$ from the minimum. Dine, Randall, and
Thomas suggested this temporary inflation generated potential as a way
to solve the
problem of moduli, but no one has come up with a plausible
way to insure that the minimum of the
potential during inflation is the same as that in the vacuum state.}.

\subsec{The Dilaton Mass and the Mechanism for SUSY Breaking}

Perhaps the simplest way to break SUSY and give the dilaton a mass is the one
described in the previous section.   The moduli are frozen by supersymmetric
dynamics at a high scale, and we
assume another gaugino condensate at a scale $e^{-{c\over 3}}M_P$.
Since the other moduli are assumed to be frozen out at the higher
scale $e^{-{b\over 3}}M_P$, this would give rise to a dilaton potential
\eqn\dilpot{V_D = e^{-c(S + S^*)}[K^{S S^*}\vert c - \partial_S K\vert^2 -3]}
Note that the minimum of this potential is not the same as that of $V_I$
defined in the previous footnote.
If we accept the assertion\ref\banksdine{T.Banks, M.Dine, Rutgers preprint
RU-50-94  hep-th@xxx.lanl.gov - 9406132, {\it To be published in Physical
Review D}.}
that we do not currently
know how to calculate $K$ for the relevant values of $S$, then it is not
implausible
that $V_D$ has a stable minimum with zero cosmological constant (we disregard
the fine tuning of the cosmological constant).  This can only happen if SUSY
is spontaneously broken, and hypothetical supersymmetric minima of $V_D$
would not be stationary points of the equations of motion after inflation.

Variations on this scheme are possible in which dynamics more complicated than
gaugino condensation generate the dilaton potential.  All one needs is a
supersymmetric gauge theory which in the absence of the dilaton ({\it i.e.}
when the coupling constant is a constant), generates a nonvanishing,
nonperturbative superpotential, and is characterized by a single energy scale.
All such theories give rise to what were referred to in \bkn as
Nonrenormalizable Hidden Sector (NRHS) Models.  In such models SUSY breaking
will be communicated to the standard model by gravitational scale dynamics,
and the nonperturbative scale $e^{- {c Re S_0 \over 3}}M_P$ must be of order
$10^{13.5}~ GeV$.
As a consequence, the dilaton will get a mass of order $1 ~ TeV$ and the
cosmology of the model will suffer from the PPP.
One is led to consider the possibility of low energy SUSY breaking.

\subsec{A Digression on the Cosmological Constant in Theories With Low Energy
SUSY breaking}

One serious problem with low energy breaking of SUSY
is the cosmological constant.
In NRHS models the cancellation of the cosmological constant is \lq\lq
natural
in order of magnitude'' \bkn.
That is, the two terms which must cancel in order
to give a zero cosmological constant naturally
 have the same order of magnitude.
This is because the breaking of R-symmetry responsible for the negative
term in the potential, is the trigger for the breaking of SUSY which gives
rise to the positive term.  The latter is a gravitational strength reaction
to the former.

In theories with low energy SUSY breaking, this is not the case.  The breaking
of SUSY is a flat space effect and the negative term in the supergravity
potential is nominally subleading by two powers of the low energy scale over
the Planck mass.  Conventionally this is ``fixed up'' by adding an appropriate
constant to the superpotential, but this is a very suspicious procedure
in string theory.  According to the rules of perturbative string theory
the constant term in the superpotential is quantized in units of the cube of
the inverse compactification radius, which is itself close to the string scale.
All other terms in the superpotential are field dependent.

In the present context, in which the low energy theory contains a dilaton
in addition to the low energy degrees of freedom which break SUSY, this problem
may be solved.  Let us suppose that the low energy theory has a nonperturbative
scale $\mu$, given in terms of the dilaton by $\mu = e^{- S\over b} M_S$. $\mu$
is the scale of dynamical SUSY breaking.
The effective action for the Goldstino has the
form
\eqn\goldstino{{\cal S}_{goldstino} = \int d^4 \theta K(X,X^* ) + \mu^2 \int
d^2 \theta X+ h.c.}
The Kahler potential is chosen so that the scalar partners of the Goldstino all
have mass of order $\mu$.  It has the form $XX^* k({X\over\mu},{X^*
\over\mu})$.   In the absence of a dilaton, this lagrangian would give a
positive cosmological constant of order $\mu^4$ when plugged into the
supergravity formula
for the potential.  The negative term in the formula is smaller than the
positive term by a factor of order ${\mu^2 \over M_P^2}$.  In the presence of a
dilaton coupled to $X$ via the $S$ dependence of $\mu$ however, we are free to
add a purely $S$ dependent term to the superpotential.  This would come from,
for example, gaugino condensation in some strongly coupled gauge theory with a
scale higher than $\mu$.

The dilaton will also control the coupling of this second gauge theory.  If
we assume that the second theory has a beta function one and a half times
as large as that of the low energy, SUSY breaking theory, then the (positive
and negative) contributions to the dilaton potential from the more
strongly coupled theory will be of the same order of magnitude as the
SUSY breaking $F$ term.  The cancellation of the cosmological constant
will no longer require the introduction of a constant superpotential with
unexplained (and in string theory, unexplainable) order of magnitude.
The $3\over 2$ ratio between beta functions that is required in this approach
could naturally arise from group theory.  The search for pairs of theories,
with the required properties might be an interesting constraint on string
vacua.

We emphasize that our considerations do not explain the precise fine tuning
of the cosmological constant, but are only order of magnitude predictions.
However, models of low energy SUSY breaking lead to an interesting range
of dilaton masses.  Depending on the mechanism for transmitting SUSY
breaking to the standard model, the scale $\mu$ can range from $1000~ TeV$
to $1~ TeV$.  This leads to dilaton masses between $100~ eV$ and $10^{-4}~ eV$.
The lower end of this range is of considerable interest.
General renormalization group arguments\foot{These arguments are due to
L.Susskind, and have not been published.} suggest that a local field theoretic
explanation of the vanishing of the cosmological constant could only make
sense if there existed a scalar with mass
 less than or equal to (the fourth root of
)the observational bound on the cosmological constant, or about $10^{-2}~ eV$.
This could arise in the present context from
 a SUSY breaking scale less than about $10~ TeV$.
While we have not demonstrated a mechanism by which such a light dilaton
could cancel the cosmological constant, at least the possibility is not ruled
out by general renormalization group arguments.

\subsec{After the Digression, The Problem Remains}

Unfortunately, the entire range of dilaton masses compatible with low energy
SUSY breaking appears to be ruled out by conventional cosmology.
In this mass range, the dilaton lifetime is longer than the age of the universe
and it leads to a postmodern Polonyi problem.  We can try to resolve this
problem by invoking the ``savior'' field of \cohen , although we would
now have to complicate the theory in order to explain the origin of a savior
field with the right mass.  Now the nuclei
produced in the aftermath of savior decay will not be destroyed by the
subsequent decay of dilatons.  However, the dilatons will come to dominate
the energy density of the universe shortly after nucleosynthesis, at an energy
scale of order $.1~ MeV$.  They would be a form of dark matter.
In such a cosmology the ratio between dark matter and radiation
energy densities at the present epoch would be $10^{8}$, while measurements
of the present temperature and density of the universe bound this ratio
by something of order $10^4$.  One would have to invoke the generation of
large amounts of entropy after nucleosynthesis to make the model consistent
with this observation, but that would also dilute the baryon content
of the universe, and would again make the model inconsistent with the data.

Indeed, the ``savior'' idea seems to run into either the nucleodestruction
problem or the above dark matter domination problem, for any values
of the moduli masses.  Nucleodestruction will be a problem if the
mass is significantly higher than an $MeV$.  Particles with mass lighter
than an $MeV$ and gravitational couplings, will have lifetimes of order
$10^{24}$ seconds or more.  This is seven orders of magnitude longer than
the age of the universe.  Such particles will dominate the energy density
at the present era to an extent ruled out by
our knowledge of the dark matter content of the universe.

Intermediate Scale inflation cannot remove a light dilaton from the universe
either, since the weak scale is much larger than the dilaton mass.
Thus, the RT solution to the cosmological moduli problem is an argument
in favor of hidden sector models for SUSY breaking.

\newsec{A Speculative Proposal}

All the models that we have considered were based on the standard
equations of homogeneous isotropic cosmology.
We would now like to show that in the context of modular physics,
the standard assumptions of homogeneity and isotropy are untenable.
The moduli may or may not
 dominate the energy density of the universe just after
inflation.  However, they certainly dominate it from about
the time that the energy density is $M^6 \over M_P^2$ (the moduli masses
are $\sim {M^3 \over M_P^2}$) until they decay.  During this period
the universe is matter dominated, and fluctuations can grow.
${\delta\rho\over\rho}$ will grow like the scale factor $R$ on all scales
inside the horizon.  It would not be sensible to take the initial value
of ${\delta\rho\over\rho}$ in modular energy to be less than the $10^{-5}$
value of primordial fluctuations.  This means that modular inhomogeneities
will go nonlinear when the scale factor has increased by a factor of
$10^5$ from its value when the universe became dominated by moduli.
At this time, the energy density will have decreased by a factor of $10^{-15}$.
On the other hand, the energy density at the time of modular decay,
is $m_{mod}^6 \over M_P^2$, which is a factor of $({m_{mod}\over M_P})^4$
times the energy density when the moduli begin to dominate.  This is less than
$10^{-64}$.  Thus the modular energy density fluctuations go nonlinear
long before moduli decay.  We have been, for the most part, conservative
in these estimates.  In fact, the moduli could dominate the energy
density right after inflation, as they would in models in which the moduli
themselves are the inflatons.  Furthermore, in models with low energy SUSY
breaking, the modular lifetime is longer than the age of the universe, and
modular energy density surely goes nonlinear before the moduli decay.
The only place in which we may have made an overestimate, is in
our ``sensible'' estimate of the initial inhomogeneity in the moduli.
However, the discrepancy between the scale at which moduli gravitationally
collapse, and that at which they decay is so large that we do not believe
that initial conditions could substantially alter our qualitative conclusion.

All discussions of modular cosmology must face up to the prediction
of a very early stage of the formation of collapsed objects.
We emphasize however that the scale of these inhomogeneities is extremely
small, and that if we successfuly get rid of moduli before nucleosynthesis,
they will have no effect whatsoever on large scale measurements of
the structure of the microwave background.

We have only the most preliminary remarks to make about what appears
to be a very complicated dynamical problem.  The two most likely
results of gravitational collapse of moduli are the formation
of stable {\it modular stars}, and the formation of black holes.
The formation of modular stars could possibly alleviate the problem
of modular domination of the universe.  The modular field strengths
inside such objects could well be very large,  and lead to a substantial
enhancement of the decay probability of modular matter into ordinary
particles.  We may expect moduli to have (for example) couplings to
photons of the form ${\phi\over M_P} F_{\mu\nu}^2$.  In the core of
a modular star, the modular field amplitude $\phi$ might be much larger
than the Planck scale, enhancing the coupling of this field to photons.
A modular star could well explode into a burst of photons,
quarks and leptons, soon after it forms.

Stable gravitationally bound configurations of moduli may not exist,
or may not be the fate of all large density fluctuations.  An alternative
would be collapse towards a black hole.  However, the work of \ref\nobh{
A.G. Polnarev, M.Yu.Khlopov, {\it Sov. Phys. Usp.}{\bf 28},(1985),213;
B.J.Carr, J.H.Gilbert, J.E.Lidsey, FermiLab Pub 94/109-A, astro-ph-9405027 .}
suggests that if the spectral index of primordial density fluctuations
is near one (as it is expected to be if the fluctuations originate from
inflation), then very few black holes will form.  We do not understand
what the evolution of the system would be if local analysis were to
preclude the existence of localized stationary solutions of the
coupled gravity-moduli system, while the global analysis of\nobh
rules out the formation of black holes.

Thus, nonlinear gravitational effects in the modular medium might provide
a mechanism for solving the modular dominance problem, but our
understanding of this complicated nonlinear regime is sketchy.
It seems clear then that all conclusions about the viability of
a cosmology which includes moduli must await the resolution of
the complicated dynamical problem of modular collapse.

\newsec{Summary and Conclusions}

In summary, our investigation of the modular cosmological problem
has taught us a number of interesting things about cosmology
and SUSY breaking.  The Intermediate Scale Inflation scenario
of Randall and Thomas seems to solve the problem if SUSY is broken
by a Nonrenormalizable Hidden Sector mechanism.  Attempts to solve
the problem by attributing moduli masses to SUSY preserving high scale
dynamics led to a number of interesting conclusions: Generic supersymmetric
stationary points of the effective potential are not stationary points
of the equations of cosmology in a postinflationary universe.  Inflation
can create large long lived smooth regions of the universe only if the
cosmological constant is greater than or equal to zero, which means
that either SUSY is broken or R-symmetry is preserved\foot{We use this
symmetry statement as a synonym for the vanishing of the superpotential
since without R-symmetry the superpotential could only vanish by
fine tuning.}.  R symmetry preservation is a very strong constraint
on the strongly coupled dynamics which is supposed to give mass to the
moduli.  If, at a generic point in string moduli space, the strongly
coupled sector is a pure gauge theory, then R symmetry can only be
preserved at special points in moduli space where massless chiral
multiplets with hidden sector gauge charge appear.  There are no known
points in moduli space where this happens.
Even if one could be found, strongly
coupled SUSY preserving dynamics cannot give mass to the dilaton.

The Postmodern Polonyi Problem posed by string theory moduli seems to us
to be a serious but potentially exciting crisis for string theory.
Although attempts to solve this problem have led to a number of
interesting ideas about string vacuum selection, SUSY breaking,
and the notion of stability in postinflationary cosmology, the only
acceptable solution in sight is Intermediate Scale Inflation (ISI).
It is important to investigate the consequences of this scenario
and to search for its origins in an explicit string vacuum solution.

We note however that extant attempts to resolve the PPP have
completely ignored the gravitational collapse of modular energy
density fluctuations which we demonstrated to be important in the very
early universe (according to string theory).  It is not clear whether
the dynamics during this era eliminates moduli without ISI, or whether
it instead leads to further problems that cannot be solved even by ISI.
Clearly, closer investigation of this primordial matter dominated
era is called for.  We hope to return to it in a future paper.

Finally, we wish to emphasize, that although we have posed the PPP in the
context of inflationary cosmology, it is equally serious in any Big Bang
cosmology.  Inflation with a Hubble constant higher than the weak
scale gives us specific predictions about the initial conditions of
moduli fields, but any model in which the energy density is of order
$(10^{11}\quad GeV)^4$ at some period in cosmic history will also predict
that the moduli start the conventional Robertson-Walker era displaced
from the minimum of their potential.  The Postmodern Polonyi Problem
cannot be evaded by rejecting inflation.

\vfill\eject
\centerline{\bf Acknowledgements}

We would like to thank our colleagues, G. Moore and S. Shenker, who
collaborated with us on research closely related to this paper, for many
important conversations about the cosmology of moduli.  T.B. also thanks
P.Binetruy, A. Cohen, M.K. Gaillard,
A. Nelson, J.Polchinski, L.Randall and S.Thomas for sharing their insights.
Much of this paper was written while T.B. was on sabbbatical at the
Ecole Normale Superieure in Paris, the University of Rome II, {\it Tor
Vergata}, and the Weizmann Institute of Science.  He would like to thank
the staff and faculty of these institutions for their hospitality. In
particular, he acknowledges the support of the Varon Visiting Professorship
while at the Weizmann Insitute.  T.B. and P.S. both acknowledge support
by the J.S. Guggenheim Memorial Foundation.   Finally, P.S. would like
to thank the members of the Rutgers High Energy Theory Group and
the Institute for Advanced Study in Princeton for their hospitality.

\listrefs
\end